\documentclass[
  journal=pasa,
  manuscript=research-paper, 
  year=2020,
  volume=37,
]{cup-journal}

\usepackage{microtype,siunitx,booktabs,amsmath}
\usepackage{aas_macros}
\usepackage{hyperref}
\usepackage{xtab,booktabs}
\hypersetup{colorlinks,citecolor=blue,linkcolor=blue,urlcolor=blue}
\newcommand{\ps}[0]{\href{https://github.com/NickSwainston/pulsar_spectra}{\texttt{pulsar\_spectra}}}

\usepackage{listings}
\definecolor{codegreen}{rgb}{0,0.6,0}
\definecolor{codered}{rgb}{0.6,0,0}
\definecolor{codeblue}{rgb}{0,0,0.6}
\definecolor{codegray}{rgb}{0.5,0.5,0.5}
\definecolor{codepurple}{rgb}{0.58,0,0.82}
\definecolor{codeorange}{RGB}{200, 112, 60}
\definecolor{backcolour}{rgb}{0.95,0.95,0.95}

\lstdefinestyle{mystyle}{
    backgroundcolor=\color{backcolour},   
    commentstyle=\color{codegreen},
    keywordstyle=\color{codepurple},
    numberstyle=\tiny\color{codegray},
    stringstyle=\color{codeorange},
    basicstyle=\ttfamily\footnotesize,
    breakatwhitespace=false,         
    breaklines=true,                 
    captionpos=b,                    
    keepspaces=true,                 
    numbers=left,                    
    numbersep=5pt,                  
    showspaces=false,                
    showstringspaces=false,
    showtabs=false,                  
    tabsize=4
}

\title{pulsar\_spectra: A pulsar flux density catalogue and spectrum fitting repository}

\author{N. A. Swainston}
\affiliation{International Centre for Radio Astronomy Research, Curtin University, Bentley, WA6102, Australia}
\email[N. A. Swainston]{nicholas.swainston@curtin.edu.au}

\author{C. P. Lee}
\affiliation{Department of Physics and Astronomy, Curtin University, Bentley, WA6102, Australia}

\author{S. J. McSweeney}
\affiliation{International Centre for Radio Astronomy Research, Curtin University, Bentley, WA6102, Australia}

\author{N. D. R. Bhat}
\affiliation{International Centre for Radio Astronomy Research, Curtin University, Bentley, WA6102, Australia}

\received {21-Jun-2022}
\revised  {01-Aug-2022}
\accepted {27-Sep-2022}
\published{DD-MMM-YYYY}


\begin{document}

\begin{abstract}
We present the \ps{} software repository, an open-source pulsar flux density catalogue and automated spectral fitting software that finds the best spectral model and produces publication-quality plots. The {\sc python}-based software includes features that enable users in the astronomical community to add newly published spectral measurements to the catalogue as they become available.
The spectral fitting software is an implementation of the method described in \cite{Jankowski2018} which uses robust statistical methods to decide on the best-fitting model for individual pulsar spectra. 
\ps{} is motivated by the need for a centralised repository for pulsar flux density measurements to make published measurements more accessible to the astronomical community and provide a suite of tools for measuring spectra.
\end{abstract}

\section{INTRODUCTION}
Although pulsars were discovered in 1967 \citep{hewish68}, the exact mechanism by which they emit electromagnetic radiation is far from being understood. The pulsar emission mechanism is often described using models that include a plasma-filled magnetosphere that co-rotates with the pulsar \citep{Goldreich1969}. However, an emission mechanism that can successfully explain all pulsar emission characteristics, including spectral turn-over and nulling, has not yet been advanced. Radio spectra of pulsars provide important clues, but accurate spectral data are lacking for the majority of pulsars.


Measurements of flux densities are important for inferring pulsar radio luminosities and energetics, as well as for detailed spectral analysis. However, producing accurate pulsar flux densities is challenging for several reasons. Firstly, pulsars scintillate due to the interstellar medium, which can cause the apparent flux density to fluctuate from a factor of two to an order of magnitude, particularly at low frequencies, depending on their Galactic latitude and the choice of instrumental parameters (e.g. observing bandwidth and time duration), on time scales up to weeks and months \citep[e.g.][]{Swainston2021, Bhat2018, Bell2016}. Obtaining flux measurements, therefore, requires observing campaigns much longer than the scintillation time scale. Secondly, flux density measurements require extensive knowledge of the telescope to account for antenna temperature, beam shape, etc. Furthermore, in the case of pulsars with severely broadened pulse profiles (due to temporal broadening resulting from multipath scattering), reliable measurements may require imaging rather than time-domain techniques. Despite these difficulties, many accurate flux density measurements have been taken over the last several decades \citep[e.g.][]{Izvekova1981, Taylor1993, Lorimer1995, Malofeev2000, Hobbs2004, Bilous2016, Han2016, Johnston2018, Jankowski2019, Sanidas2019}. However, there is currently no catalogue designed to record pulsar flux density measurements at arbitrary frequencies. Researchers are obliged to do extensive literature reviews, find the publications that contain flux density measurements of pulsars they are interested in and extract the information from them. This exercise is a time-consuming task that is prone to error.

There is no complete theoretical model for pulsar spectra. For this reason, we use several empirical
models as no single model can accurately fit the variety of pulsars' spectra. Deciding which models to use and which is best for each pulsar requires sophisticated statistical techniques. \cite{Jankowski2018} has detailed a method for deciding on the best model using the Akaike information criterion (AIC), which measures the information each model retains without overfitting. This is applied to various spectral models used throughout the literature (see \S \ref{sec:models}). Their choice of method and implementation can result in different results compared to other researchers with the same data.

We have implemented the methodology of \cite{Jankowski2018} in \ps{}\footnote{\url{https://github.com/NickSwainston/pulsar_spectra}}, a fully-featured spectral fitting {\sc python} software package.
It includes all five models listed above and can be easily extended to include others.
\ps{} also contains a catalogue of flux density measurements from several publications.
\ps{} is open-source; researchers can upload measurements from new publications into the catalogue, which are then available to all. The software has already been used in \cite{Lee2022}.

The remainder of this paper is organised as follows. We will first explain the implementation of the catalogue and its benefits in \S \ref{sec:catalogue}. Then in \S \ref{sec:spectral_fit}, we discuss the method used to determine the best spectral fit. Finally, in \S \ref{sec: software} we demonstrate how to use the software.

\section{CATALOGUE} \label{sec:catalogue}
There is currently no pulsar catalogue exclusively for flux density measurements. The ATNF pulsar catalogue \citep[version v1.67;][]{manchester05} does maintain a large collection of published flux densities, but the frequency at which they are measured is not always accurate.
Flux densities measured at frequencies other than a set of pre-selected frequencies (currently 26, ranging from \SI{30}{\MHz} to \SI{150}{\GHz}) are typically recorded at the nearest listed frequency, which can lead to inaccurate spectral fits if the frequency and flux density values in the catalogue are taken at face value. Moreover, the catalogue's design prohibits multiple measurements at the same or similar frequencies, with only the most recent measurement recorded at any given (approximate) frequency.

Researchers are currently obliged to conduct their own extensive literature reviews to find the publications that contain flux density measurements of pulsars they are interested in and extract the relevant information from them. This is a time-consuming task and, because there is no central place to store this information, it is duplicated effort by each researcher.

To allow researchers to acquire accurate flux density measurements with minimal effort, we have created an open-source catalogue within \ps{}. The catalogue consists of dictionaries in the form of a YAML file for each included publication. These YAML files contain all the flux density measurements and their uncertainty in mJy and their frequencies in MHz for each pulsar. If the original authors gave no flux density uncertainty, we assumed a conservative relative uncertainty of 50\%, following earlier work \citep{Sieber1973, Jankowski2018}. This catalogue can easily be collected using a {\sc python} function and combined with new results to produce a pulsar spectral fit (as demonstrated in \S \ref{sec: software}).

\begin{figure*}[htbp]
\centering
\includegraphics[height=0.9\textheight]{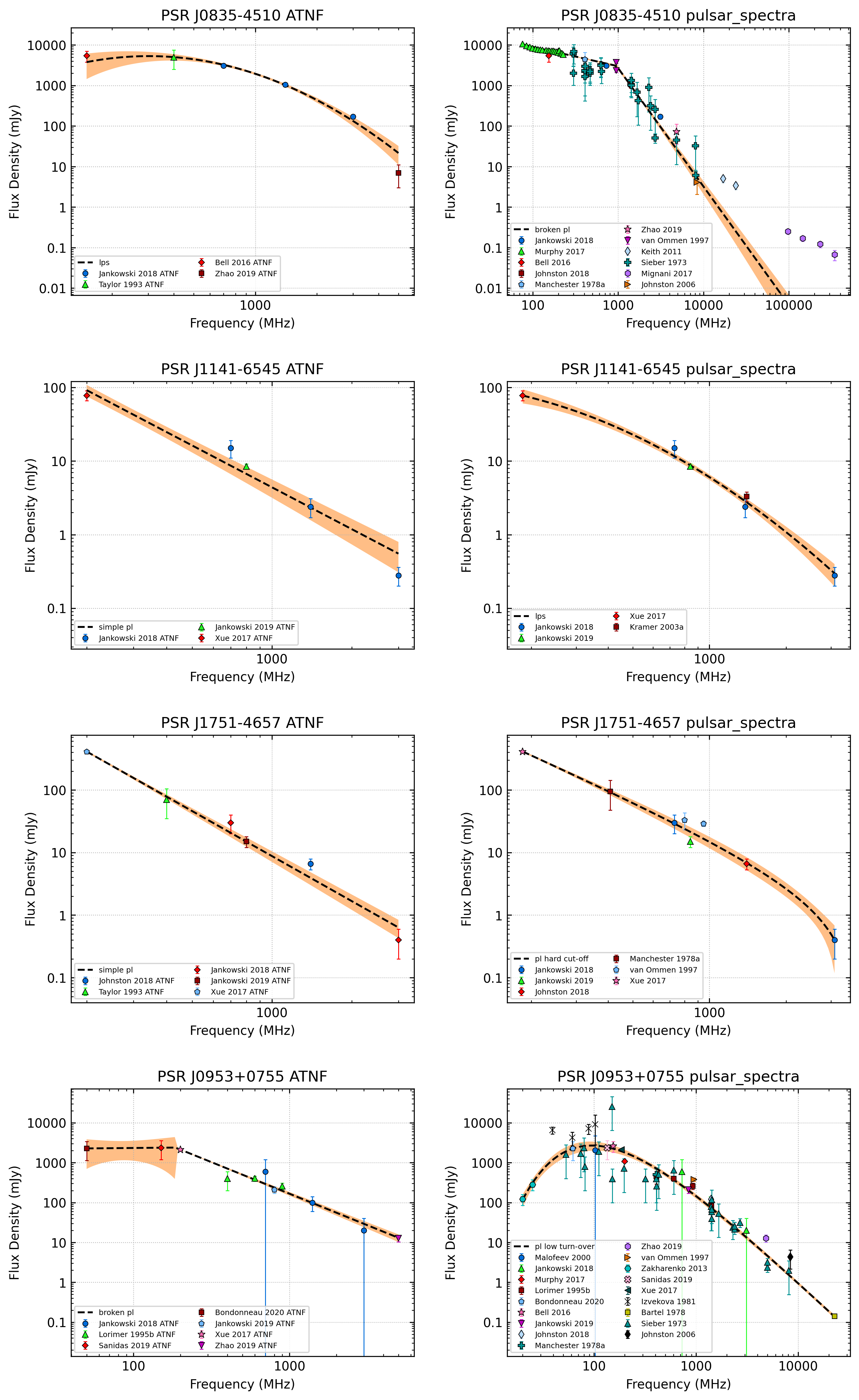}
\caption{\textit{Left}: A spectral fit using only the flux density values from the ATNF pulsar catalogue. \textit{Right}: A spectral fit using only the flux density values from the \ps{} catalogue. As demonstrated through these examples, the \ps{} catalogue is able to accommodate  many more flux density values than those available in the ATNF pulsar catalogue, and hence can yield more accurate spectral fits.}
\label{fig:atnf}

\end{figure*}

When using the flux density values from the ATNF to fit a pulsar, most researchers know that the values are inaccurate due to the select frequencies of the ATNF (see the left plots in Figure \ref{fig:atnf}). They would then extract the publications' true flux density and frequency values to create a more accurate fit. With \ps{} this is rarely required as our catalogue can use any frequency value (see the right plots in Figure \ref{fig:atnf}). This frequency flexibility also allows us to include papers, such as \cite{Murphy2017} and \cite{Johnston2006}, which have flux density measurements at many frequencies. The four pulsars in Figure \ref{fig:atnf} are examples of a different and more accurate fit using \ps{} without manually extracting values from the publications.

\subsection{Currently included and future publications}

The catalogue is designed to be a community-maintained, open-source catalogue that prevents duplicated effort.
We have currently added 34 publications to the database, which are shown in Table \ref{table:catalogue_refs}. This is not a complete list of pulsar flux density publications and is likely to favour Southern-sky (declination $\delta < 0$) pulsars. 
As researchers use this catalogue, they can add new flux density measurements (or historical ones that are not already included), which can then be made available for other researchers.

To make it easier for other researchers to include new publications in the catalogue, we have created a script to convert a simple CSV into the required YAML format that the catalogue requires, as explained in \href{https://pulsar-spectra.readthedocs.io/en/latest/catalogue.html\#adding-to-the-catalogue}{our documentation}\footnote{\url{https://pulsar-spectra.readthedocs.io/en/latest/catalogue.html\#adding-to-the-catalogue}}. They can use this to make a pull request, and this publication will be included in the next release of \ps. We will keep an up-to-date \href{https://pulsar-spectra.readthedocs.io/en/latest/catalogue.html\#papers-included-in-our-catalogue}{table of publications}\footnote{\url{https://pulsar-spectra.readthedocs.io/en/latest/catalogue.html\#papers-included-in-our-catalogue}} to make it easier to cite the data and to encourage authors to upload their own published results.

\tablecaption{The publications included in version 2.0 of \ps. For an up to date table see the  \href{https://pulsar-spectra.readthedocs.io/en/latest/catalogue.html\#papers-included-in-our-catalogue}{documentation}}
\tablefirsthead{\midrule Publication & \# Pulsars & $\nu$ (MHz) \\ \midrule \midrule}

\tablehead{
    \multicolumn{3}{c}
    {{\bfseries  Continued from previous column}} \\
    \toprule
    First&\multicolumn{1}{c}{Name}\\ \midrule
}

\tabletail{\midrule \multicolumn{2}{r}{{Continued on next column}} \\ \midrule}
\begin{xtabular}{lrl}
ATNF pulsar catalogue & 2827 & 40-150000 \\
\cite{Sieber1973} & 27 & 38-10690 \\
\cite{Bartel1978} & 18 & 14800-22700 \\
\cite{Manchester1978a} & 224 & 408-408 \\
\cite{Izvekova1981} & 86 & 39-102 \\
\cite{Dewey1985} & 34 & 390-390 \\
\cite{McConnell1991} & 4 & 610-610 \\
\cite{Johnston1992} & 100 & 640-1500 \\
\cite{Wolszczan1992} & 1 & 430-1400 \\
\cite{Johnston1993} & 1 & 430-2360 \\
\cite{Manchester1993} & 1 & 640-640 \\
\cite{Taylor1993} & 639 & 400-1400 \\
\cite{Camilo1995} & 29 & 430-430 \\
\cite{Lundgren1995} & 1 & 430-1400 \\
\cite{Nicastro1995} & 1 & 400-1400 \\
\cite{Qiao1995} & 61 & 600-1500 \\
\cite{Robinson1995} & 2 & 436-640 \\
\cite{Lorimer1995} & 280 & 408-1606 \\
\cite{Manchester1996} & 55 & 436-436 \\
\cite{Zepka1996} & 1 & 430-1400 \\
\cite{vanOmmen1997} & 82 & 800-960 \\
\cite{Malofeev2000} & 212 & 102-102 \\
\cite{Kramer2003a} & 200 & 1400-1400 \\
\cite{Hobbs2004a} & 453 & 1400-1400 \\
\cite{Karastergiou2005} & 48 & 1400-3100 \\
\cite{Johnston2006} & 31 & 8400-8400 \\
\cite{Lorimer2006} & 142 & 1400-1400 \\
\cite{Kijak2007} & 11 & 325-1060 \\
\cite{Stappers2008} & 13 & 147-147 \\
\cite{Bates2011} & 34 & 1400-6500 \\
\cite{Keith2011} & 9 & 17000-24000 \\
\cite{Kijak2011} & 15 & 610-4850 \\
\cite{Zakharenko2013} & 40 & 20-25 \\
\cite{Dembska2014} & 19 & 610-8450 \\
\cite{Dai2015} & 24 & 730-3100 \\
\cite{Stovall2015} & 36 & 35-79 \\
\cite{Basu2016} & 1 & 325-1280 \\
\cite{Bell2016} & 17 & 154-154 \\
\cite{Bilous2016} & 158 & 149-149 \\
\cite{Han2016} & 204 & 1274-1466 \\
\cite{Kijak2017} & 12 & 325-610 \\
\cite{Mignani2017} & 1 & 97500-343500 \\
\cite{Murphy2017} & 60 & 76-227 \\
\cite{Xue2017} & 50 & 185-185 \\
\cite{Jankowski2018} & 441 & 728-3100 \\
\cite{Johnston2018} & 586 & 1400-1400 \\
\cite{Jankowski2019} & 205 & 843-843 \\
\cite{Sanidas2019} & 290 & 135-135 \\
\cite{Xie2019} & 32 & 300-3000 \\
\cite{Zhao2019} & 71 & 4820-5124 \\
\cite{Bilous2020} & 43 & 53-63 \\
\cite{Bondonneau2020} & 64 & 53-65 \\
\cite{McEwen2020} & 670 & 350-350 \\
\cite{Han2021} & 201 & 1250-1250 \\
\cite{Johnston2021} & 44 & 1400-1400 \\
\hline
\label{table:catalogue_refs}
\end{xtabular}

When comparing the current state of the catalogue to the ATNF catalogue (see Table \ref{table:catalogue_compare}), the \ps{} catalogue already contains a larger sample of pulsar flux density measurements. We hope that the open-source nature of our catalogue and the ease of uploading new publications' flux density measurements will allow our catalogue to grow rapidly in the coming years.

\begin{table}
\begin{center}
\begin{tabular}{lr}
\hline
Source & \# pulsars \\
\hline \hline
Only ANTF           & 812 \\
Both catalogues     & 581 \\
Only pulsar\_spectra & 1385 \\
Neither catalogues  & 541 \\
\hline
\end{tabular}
\caption{This table compares the current progress of our catalogue compared to the ATNF. As users continue to add publications the catalogue will grow.}
\label{table:catalogue_compare}
\end{center}
\end{table}

\section{PULSAR SPECTRAL FITTING} \label{sec:spectral_fit}
Flux density measurements obtained using different telescopes are subject to different systematic errors due to the telescope's observing setup and varying levels of reliability of the calibration procedures. These systematic errors make robust modelling of spectral fits complicated. \cite{Jankowski2018} developed a method of modelling and objectively classifying spectra which are composed of disparate data from the literature, and our approach is adapted from this work. We now summarise the \cite{Jankowski2018} method used in our software.

To reduce the effect of underestimated uncertainties on outlier points in a given fit, the least-squares function is modified from the regular quadratic loss to a linear loss once the residuals exceed a pre-chosen threshold. In this way, outlier data are penalised, and as a result any measurements that are less reliable are less likely to skew the model fit. In \ps{}, we implement this using the Huber loss function, defined as
\begin{equation}
    \rho =
    \begin{cases}
    \frac{1}{2}t^2 & \mathrm{if}\:|t|<k \\[5pt]
    k|t|-\frac{1}{2}k^2 & \mathrm{if}\:|t|\geq k
    \end{cases},
\end{equation}
where $t$ is a residual (i.e. the difference between the model and the measurement) and $k$ is the threshold that defines which points are considered outliers \citep{Huber1964}. We use a value of $k = 1.345$, for which Huber has shown to be 95\% as efficient at parameter estimation as an ordinary least squares estimator operating on data from a Gaussian distribution. We hence define a robust cost function
\begin{equation}\label{eq:rcf}
    \beta=\sum_i^N
    \begin{cases}
    \frac{1}{2}\left(\frac{f_i-y_i}{\sigma_{y,i}}\right)^2 & \mathrm{if}\:\left|\frac{f_i-y_i}{\sigma_{y,i}}\right|<k \\[5pt]
    k\left|\frac{f_i-y_i}{\sigma_{y,i}}\right|-\frac{1}{2}k^2 & \mathrm{otherwise}
    \end{cases},
\end{equation}
where $f_i$ are the values of the model function at the frequencies of the measured flux densities $y_i$, and $\sigma_{y,i}$ are the corresponding uncertainties of the flux densities.

The cost function is minimised using {\sc migrad}, a robust minimisation algorithm implemented in the {\sc minuit c++} library \citep[as described in][]{James1975} which is accessible through the {\sc python} interface {\sc  iminuit}\footnote{\url{https://github.com/iminuit/iminuit}}. {\sc migrad} uses a combination of Newton steps and gradient descents to converge to a local minimum. The Estimated Distance to Minimum (EDM) is used to define a convergence criterion in terms of a specified tolerance (which must be met for the minimisation to be considered successful), which is set to a value of \num{5e-6}. The uncertainties are computed at the 1$\sigma$ level from the diagonal elements of the parameter covariance matrix using the {\sc hesse} error calculator.

\subsection{Spectral models}\label{sec:models}
We have currently implemented the five spectral models that have distinct spectral shapes from \cite{Jankowski2018}. While these five models are sufficient for describing the spectra for the vast majority of pulsars, our software is flexible enough to allow the addition of more spectral models easily, as explained \href{https://pulsar-spectra.readthedocs.io/en/latest/spectral_fit.html\#models}{here}\footnote{\url{https://pulsar-spectra.readthedocs.io/en/latest/spectral_fit.html\#models}}.

The choice of the reference frequency, $\nu_0$, in the following models can affect the spectral fits as the data closer to the reference frequency are given more weight. To make the weighting of the fit as even as possible, we select $\nu_0$ as the geometric average (average in log-space) of the minimum and maximum frequencies in the spectral fit ($\nu_\text{min}$ and $\nu_\text{max}$, respectively):
\begin{equation}
    \log_{10}\nu_0 = \frac{1}{2}\left(\log_{10}\nu_\text{min}+\log_{10}\nu_\text{max}\right).
\end{equation}
This method has also been adopted in previous work \citep[e.g.][]{Bilous2016}. We define the scaling constant in the following spectral models as $c$.

We shall now describe the spectral models and provide example pulsars that are best fit by each of these models as of \ps{} version 2.0.

\subsubsection{Simple power law}
The simple power law is linear in log-space and the most common spectral model. Some examples of this include PSRs J0034-0534 (see Figure \ref{fig:J0034-0534}), J1328-4357, and J0955-5304. The model takes the form:
\begin{equation}
    S_\nu = c \left( \frac{\nu}{\nu_0} \right)^\alpha,
\end{equation}
where $\alpha$ is the spectral index. The fit parameters are $\alpha$ and $c$.

\subsubsection{Broken power law}
The broken power law is the equivalent of two simple power laws that differ at a spectral break frequency.
Variations on this model have been applied in other fields of astrophysics, such as the smoothly broken power-law, which is characterised by an additional parameter describing the width of the transition \citep[e.g.][]{Ryde1999}.
Pulsar spectra are traditionally fit with sharply broken power laws, characterised by only four free parameters \citep[e.g.][]{Sieber1973,Murphy2017}.
Some examples of broken power-law fits include PSRs J0437-4715 (see Figure \ref{fig:atnf}), J0452-1759 and J0820-1350. The model takes the form:
\begin{equation}
    S_\nu = c\begin{cases}
            \left( \frac{\nu}{\nu_0} \right)^{\alpha_1}   & \mathrm{if}\: \nu \leq \nu_b \\[5pt]
            \left( \frac{\nu}{\nu_0} \right)^{\alpha_2} \left( \frac{\nu_b}{\nu_0} \right)^{\alpha_1-\alpha_2} & \mathrm{otherwise} \\
        \end{cases},
\end{equation}
where $\nu_b$ is the frequency of the spectral break, $\alpha_1$ the spectral index before and $\alpha_2$ the one after the break. The fit parameters are $\alpha_1$, $\alpha_2$, $\nu_b$, and $c$.

\subsubsection{Log-parabolic spectrum}
This model has been used to describe the spectra of radio galaxies \citep[e.g.][]{Baars1977} and curved pulsar spectra \citep{Bates2013, Dembska2014}. Some examples of log-parabolic spectra include PSRs J1141-6545 (see Figure \ref{fig:atnf}), J1313+0931 and J0837+0610. The model takes the form:
\begin{equation}
    \mathrm{log}_{10} S_\nu = a  \left [ \mathrm{log}_{10} \left ( \frac{\nu}{\nu_0} \right ) \right]^2 + 
                            b \, \mathrm{log}_{10} \left ( \frac{\nu}{\nu_0} \right ) + c
\end{equation}
where $a$ is the curvature parameter and $b$ is the spectral index for $a = 0$. The fit parameters are $a$, $b$, and $c$.

\subsubsection{Power law with high-frequency cut-off}
This model, also known as the hard cut-off model, is based on the coherent emission model developed by \cite{Kontorovich2013}. At high frequencies, its flux quickly trends towards zero before a cut-off frequency. An example of this is PSR J1751-4657 (see Figure \ref{fig:atnf}). The model takes the form:
\begin{equation}
    S_\nu = c\left( \frac{\nu}{\nu_0} \right)^{\alpha} \left ( 1 - \frac{\nu}{\nu_c} \right ),\qquad \nu < \nu_c,
\end{equation}
where $\alpha$ is the spectral index and $\nu_c$ is the cut-off frequency. The fit parameters are $\alpha$, $\nu_c$ and $c$.

\subsubsection{Power law with low-frequency turn-over}
This model exhibits a power law at high frequencies with a turn-over at low frequencies. The curve of this turn-over can give us clues about the nature of the pulsar emission mechanism \citep{Izvekova1981, Kijak2007}. Some examples of this include PSRs J0953+0755 (see Figure \ref{fig:atnf}), J1543+0929 and J0034-0721. The model takes the form:
\begin{equation}
    S_\nu = c\left( \frac{\nu}{\nu_0} \right)^{\alpha} \exp\left [ \frac{\alpha}{\beta} \left( \frac{\nu}{\nu_{peak}} \right)^{-\beta} \right ],
\end{equation}
where $\alpha$ is the spectral index, $\nu_{peak}$ is the turn-over frequency, and $0 < \beta \leq 2.1$ determines the smoothness of the turn-over. The fit parameters are  $\alpha$, $\nu_{peak}$, $\beta$ and $c$.


\subsection{Comparing models}

To compare five models (described in \S \ref{sec:models}), we require a comparison metric that accounts for a different number of fit parameters. We use the Akaike information criterion (AIC), which is a measure of how much information the model retains about the data without over-fitting. In other words, a model with more fit parameters is only rated better if the fit is sufficiently improved. It was implemented as
\begin{equation}\label{eqn:AIC}
    \mathrm{AIC}=2\beta_\mathrm{min} + 2K + \frac{2K(K+1)}{N-K-1},
\end{equation}
where $\beta_\mathrm{min}$ is the minimised robust cost function, $K$ is the number of free parameters, and $N$ is the number of data points in the fit. The last term is the correction for finite sample sizes, which goes to zero as the sample size gets sufficiently large.

The model which results in the lowest AIC is the most likely to be the model that most accurately describes the pulsar's spectra.

\section{HOW TO USE THE SOFTWARE} \label{sec: software}
\begin{figure}[ht]
\centering
\includegraphics[width=\textwidth]{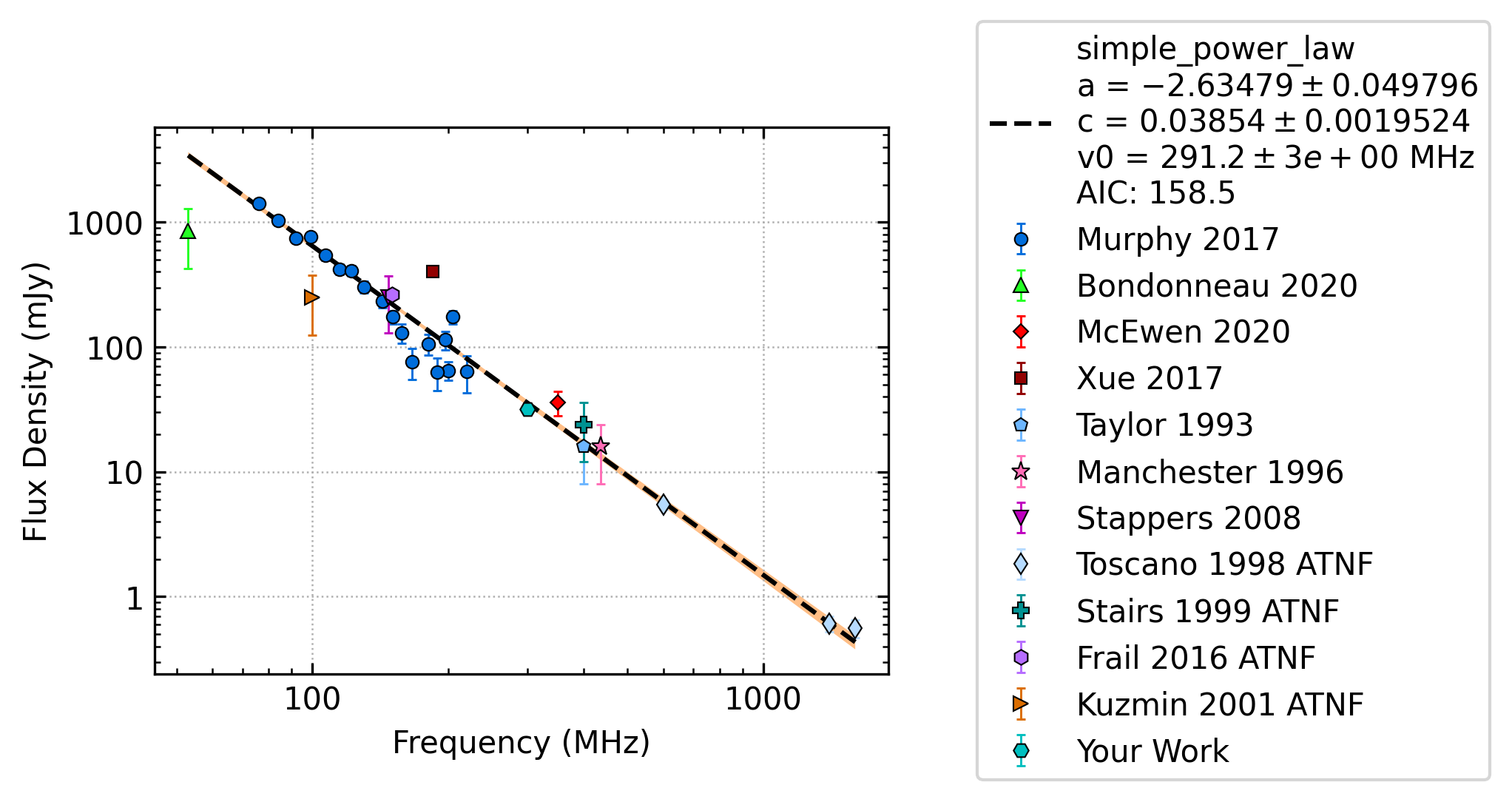}
\caption{The spectral fit to the flux density measurements of PSR J0034-0534 created using the \ps{} software and only the 11 lines of code shown in Listing \ref{list:example}.}
\label{fig:J0034-0534}
\end{figure}

\ps{} is written in {\sc python} and is easily installed using \texttt{pip install pulsar\_spectra}. The complete documentation of the code can be found \href{https://pulsar-spectra.readthedocs.io/en/latest/index.html}{here}\footnote{\url{https://pulsar-spectra.readthedocs.io/en/latest/index.html}}. To demonstrate how easy it is to use, we present a simple example for PSR J0034-0534. The code in Listing \ref{list:example} shows how to add a custom set of flux density measurements to those already in the catalogue and find the best spectra fit. The result is shown in Figure \ref{fig:J0034-0534}.

\begin{lstlisting}[language=Python, label={list:example}, caption=An example of how to add new flux density measurement to the values from the literature and find the best spectra fit.]
from pulsar_spectra.catalogue import collect_catalogue_fluxes
from pulsar_spectra.spectral_fit import find_best_spectral_fit

cat_list = collect_catalogue_fluxes()
pulsar = 'J0034-0534'
freqs, fluxs, flux_errs, refs = cat_list[pulsar]
freqs += [300.]
fluxs += [32.]
flux_errs += [3.]
refs += ["Your Work"]
find_best_spectral_fit(pulsar, freqs, fluxs, flux_errs, refs, plot_best=True)
\end{lstlisting}

\subsection{Future plans}
One of the benefits of having an open source repository is that we can continue to add models and features as pulsar spectral theory improves. One example is the implicit assumption that the reported flux density can be treated as the flux density at one specific frequency (usually the central frequency of the observing band). Such an approximation becomes increasingly inaccurate for wider and wider bandwidths. We will expand the catalogue's database to include the bandwidth, of all measurements where this information has been recorded, and expand our equations to model the integrated flux across the band.

\section{Summary}
We have introduced \ps{}, a software repository to make pulsar flux density measurements more accessible to the community and make the investigation of pulsar spectra easier by automating pulsar spectral analyses via several standard functional forms.
The open-source pulsar flux density catalogue is designed to be extendable, allowing the community to include new publications in the catalogue and cite the work of others. The analysis of spectra for a large body of pulsars can provide valuable clues to the nature of the pulsar emission mechanism and will help refine the knowledge of the detectable pulsar population.

\begin{acknowledgement}
This repository made use of the
VizieR catalogue access tool, CDS, Strasbourg, France;
{\sc matplotlib}, a {\sc python} library for publication quality graphics \citep{Hunter2007};
{\sc pandas}, a data analysis and manipulation {\sc python} module \citep{Mckinney2010};
{\sc psrqpy}, a {\sc python} module that provides an interface for querying the ATNF pulsar catalogue \citep{Pitkin2018};
{\sc iminuit}, a fast {\sc python} minimiser \citep{Hans2020};
and {\sc numpy} \citep{vanderwalt2011}.
We thank the anonymous referee for their useful suggestions for improvement of this manuscript and the repository.
\end{acknowledgement}
\bibliography{references_edited}


\end{document}